\documentclass[aps,prl,showpacs,twocolumn,superscriptaddress]{revtex4}
\usepackage{epsfig}
\usepackage{bm}
\usepackage{amsmath}

\def\be{\begin{equation}}       \def\ee{\end{equation}}
\def\bea{\begin{eqnarray}}      \def\eea{\end{eqnarray}}

\begin{document}

\title{Magic Doping Fractions in High-Temperature Superconductors}

\author{Seiki Komiya}
\affiliation{Central Research Institute of Electric Power
Industry, Komae, Tokyo 201-8511, Japan}
\author{Han-Dong Chen}
\affiliation{Department of Physics, McCullough building, Stanford University,
Stanford, CA 94305}
\author{Shou-Cheng Zhang}
\affiliation{Department of Physics, McCullough building, Stanford University,
Stanford, CA 94305}
\author{Yoichi Ando}
\affiliation{Central Research Institute of Electric Power
Industry, Komae, Tokyo 201-8511, Japan}

\begin{abstract}

We report hole-doping dependence of the in-plane resistivity $\rho_{ab}$
in a cuprate superconductor La$_{2-x}$Sr$_x$CuO$_4$, carefully examined
using a series of high-quality single crystals. Our detailed
measurements find a tendency towards charge ordering at particular
rational hole doping fractions of 1/16, 3/32, 1/8, and 3/16. This
observation appears to suggest a specific form of charge order and
is most consistent with the recent theoretical prediction of
the checkerboard-type ordering of the Cooper pairs at rational doping
fractions $x=(2m+1)/2^n$, with integers $m$ and $n$.

\end{abstract}

\maketitle

All high-$T_c$ cuprates contain three robust phases --- the insulating
antiferromagnetic (AF) phase, the superconducting (SC) phase, and the
metallic phase --- depending on the density of charge carriers
introduced by doping. However, in some cuprate materials, there are also
other electronic phases which compete with superconductivity
\cite{SACHDEV2002,KIVELSON2003}. Determining the nature of these
competing phases is a key focus of the current research in
high-temperature superconductivity. One particularly important type of
competing phase is a charge-ordered phase in underdoped cuprates, where
the carrier density is smaller than the optimum level for
superconductivity. In the underdoped regime, the mean kinetic energy of
the carriers is low because of the small carrier density, and the
Coulomb interaction plays an important role. The Coulomb interaction
generally prefers some form of charge order, whose detailed form could
be affected by the local antiferromagnetic exchange energy as well. One
possibility is that the charges form one-dimensional (1D) stripes
\cite{KIVELSON2003,ZAANEN1989,KATO1990,WHITE1998,VOJTA1999}.
Experimentally, magnetic and lattice neutron scattering on
La$_{2-x}$Sr$_x$CuO$_4$ (LSCO) and its family compounds has been
interpreted in terms of the 1D stripe picture
\cite{KIVELSON2003,TRANQUADA1995}, although the two-dimensional (2D)
nature of the spin system has recently been emphasized in Refs.
\cite{CHRISTENSEN2004,HAYDEN2004}. Considering the presence of strong
pairing interactions in this material, Chen {\it et al.}
\cite{CHEN2002,CHEN2003,CHEN2004} have proposed a 2D checkerboard-type
ordering of the hole pairs. It offers a natural explanation of the
scanning tunneling microscopy (STM) results on
Bi$_2$Sr$_2$CaCu$_2$O$_{8+\delta}$ (BSCCO) and
Ca$_{2-x}$Na$_x$CuO$_2$Cl$_2$ (NCCOC) compounds, which show rotationally
symmetric $4a\times 4a$ charge ordering patterns
\cite{HOFFMAN2002,HOWALD2003,VERSHININ2004,McElroy2004,HANAGURI2004}.
The checkerboard state of the Cooper pairs has also been discussed in
other frameworks in the recent literature
\cite{ALTMAN2002,VOJTA2002,TESANOVIC2004,ANDERSON2004}. Furthermore, the
possibility of a Wigner crystal of single holes has also been proposed
as a competing charge ordered state at low doping
\cite{FU2004,KIM2001,ZHOU2003}. In view of the contrasting experimental
results and theoretical proposals, more systematic studies of the nature
of the charge order in the cuprates are clearly called for.

The charge ordering tendency is expected to be particularly pronounced
near certain ``magic" doping levels, where the charge modulation is
commensurate with the underlying lattice
\cite{CHEN2003,TESANOVIC2004,ANDERSON2004,FU2004}. Motivated by the
recent discussions on the stripe versus the checkerboard order, we carry
out a systematic study of the doping dependence of the resistivity, in
order to uncover the possible commensurability effects. Thanks to the
greatly improved crystal-growth technique for LSCO using floating-zone
furnaces, single crystals of LSCO of unprecedented quality have recently
become available \cite{KOMIYA2002} for a very wide doping range. The
cleanliness of the new-generation crystals has allowed, for example, to
produce 100\%-untwinned single crystals \cite{Lavrov2001}, which in turn
led to finding of novel physics in this system
\cite{ANDO2002,DUMM2003,LAVROV2002}. In this work, we systematically
measure the temperature dependence of the in-plane resistivity
$\rho_{ab}$ in a series of high-quality LSCO crystals for $x$ = 0.009 --
0.216. The raw data of $\rho_{ab}(T)$ are shown in Fig. 1 for all the
superconducting samples; note that the hole doping is changed in very
small increments (typically 1\%) here, which is necessary for analyzing
how exactly the mobility of the holes changes with their density.

\begin{figure}
\centering\epsfig{file=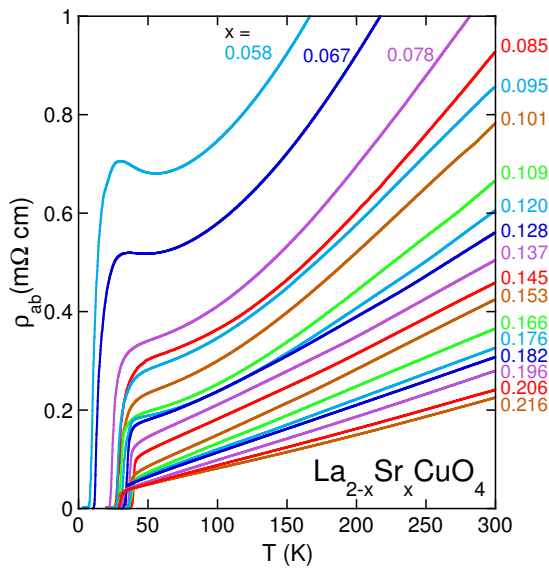,clip=1,width=0.85\linewidth,angle=0}
\caption{Temperature dependences of $\rho_{ab}$ for a series of LSCO
single crystals. The $x$ values shown are the actual Sr contents
measured by the inductively-coupled-plasma atomic-emission spectroscopy
(ICP-AES). The samples are carefully annealed to remove excess oxygens
or oxygen vacancies, so the hole density in the present samples are
essentially equal to $x$.} 
\end{figure}

Figure 2(a) shows the $x$ dependence of the inverse mobility $\mu^{-1}$,
which is equal to $ne\rho_{ab}$, for representative temperatures (the
hole density $n$ is given by $x/V$, where $V$ is the unit volume per
Cu). Also, since the absolute values of $\mu^{-1}$ are subject to
possible geometrical-factor errors (which can be up to 5\% in the case
of our measurements), in Fig. 2(b) we show
$\rho_{ab}(T)/\rho_{ab}(300{\rm K})$, which factors out such
geometrical-factor errors. One can easily see that at high temperature
the $x$ dependencies of these variables are rather smooth and
featureless, but with lowering temperature they start to ``oscillate"; a
peak at $x \simeq$ 0.13 is particularly evident. In addition, there are
three more peaks and/or shoulders, if weaker, at $x \simeq$ 0.06, 0.09,
and 0.18. This observation suggests that there are particular carrier
densities where the hole motion tends to be hindered, which weakly
enhances the resistivity at low temperature. Most naturally, such a
behavior is indicative of a ``commensurability" effect associated with
some sort of charge ordering
\cite{CHEN2003,TESANOVIC2004,ANDERSON2004,FU2004}. Remember, in usual
charge ordered systems where the Peierls transition is responsible, a
sharp increase in resistivity is observed upon charge ordering
\cite{GRUNER1988}; in the present case, where the Coulomb interaction is
likely to be responsible, the effect appears to be milder. The observed
decimal numbers (0.06, 0.09, 0.13 and 0.18) suggest that the
commensurability effect is possibly taking place at rational doping
levels 1/16, 3/32, 1/8, and 3/16. (We note that there has been some
preliminary evidence for a charge ordering tendency at $x$ = 1/16
\cite{KIM2001,ZHOU2003}). 

Given that the inverse mobility shows an $x$-dependence that is
indicative of charge ordering, one may wonder how the superconducting
transition temperature $T_c$ changes with $x$; the inset of Fig. 2(a)
shows the $x$-dependence of the zero-resistivity $T_c$ in our series of
samples. Besides a plateau-like feature for $x$ = 0.08 -- 0.12, the
$T_c$ changes rather smoothly without showing clear dips that can be
associated with the magic fractions observed in the resistivity data.
Hence, the putative charge order appears to be {\it not} particularly
destructive to superconductivity; this is rather surprising, but is
probably related to the STM observation \cite{VERSHININ2004} that the
checkerboard order shows up as a {\it precursor} to superconductivity.
In this regard, it is useful to note that most of the other experiments
concerning the checkerboard state were done below $T_c$
\cite{HOFFMAN2002,HOWALD2003,McElroy2004,HANAGURI2004}, while the
existence of the magic doping fractions is suggested in the resistivity
data above $T_c$; if the charge ordering phenomena in cuprates involve
the Cooper pairs
\cite{CHEN2003,ALTMAN2002,VOJTA2002,TESANOVIC2004,ANDERSON2004} and
those pairs are formed at $T > T_c$, it is possible that the charge
ordering, observable when the superconductivity is weakened, is
essentially of the same nature across $T_c$.

\begin{figure}
\centering\epsfig{file=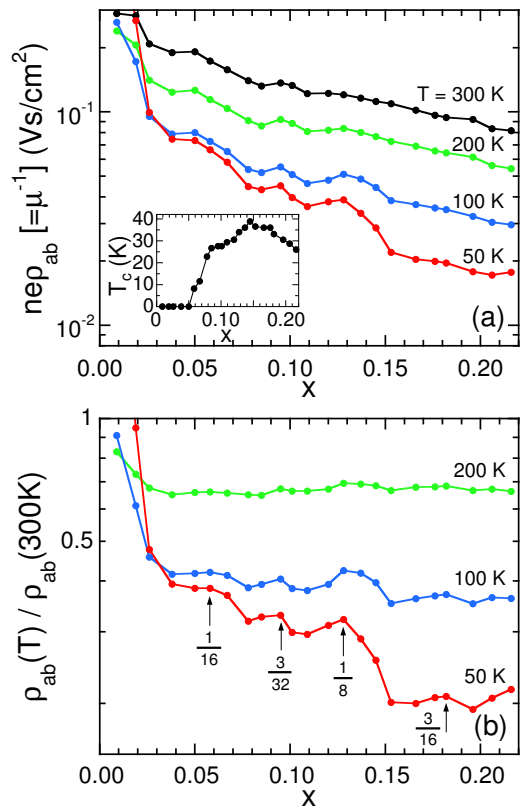,clip=1,width=0.8\linewidth,angle=0}
\caption{(a) $x$-dependence of the inverse mobility $\mu^{-1}$ 
(= $ne\rho_{ab}$) at representative temperatures. Inset shows 
the $x$-dependence of zero-resistivity $T_c$. (b) $x$-dependence 
of $\rho_{ab}(T)/\rho_{ab}(300{\rm K})$ at $T$ = 200, 100, and 50 K. 
The hole motion tends to be hindered at low temperature at 
$x \simeq$ 0.06, 0.09, 0.13, and 0.18, which corresponds to the 
``magic" doping fractions of 1/16, 3/32, 2/16, and 3/16. } 
\end{figure}

Now let us discuss the theoretical implications of our data.
The 1D stripe model predicts a particular set of ``magic" doping
fractions. The stripe model most often discussed in the literature
involves site-centered, horizontal or vertical charge stripes separated
by the AF domains at a commensurate distance of $d=pa$, where $p$ is an
integer and $a$ is the lattice constant. (For the case of $p=4$, see,
for example, Fig. 1 of Ref. \cite{TRANQUADA1995}). The holes fill
alternating sites on the charge stripe. In the stripe literature, it is
commonly assumed that the hole doping on the stripe stays fixed, while
the inter-stripe separation varies to accommodate different values of
the doping level. This simple picture predicts magic doping fractions of
$x=1/2p$, with a charge unit-cell of $2a \times pa$. 

On the other hand, the 2D checkerboard-type order generally leads to a
different set of magic doping fractions. Stimulated by the checkerboard
orders observed by STM
\cite{HOFFMAN2002,HOWALD2003,VERSHININ2004,McElroy2004,HANAGURI2004}, a
global phase diagram of cuprate superconductors has been theoretically
proposed and numerically analyzed \cite{CHEN2002,CHEN2003,CHEN2004},
where the zero-temperature phase diagram was examined in the
two-dimensional parameter space of chemical potential versus the ratio
of the hole kinetic energy over the Coulomb interaction, within the
framework of the $SO(5)$ theory. Most intriguingly, this theory
predicts, besides the AF and SC states, checkerboard-type ordering of
the Cooper pairs at magic rational doping fractions $(2m+1)/2^n$, where
$m$ and $n$ are integers \cite{CHEN2003}. A hierarchy construction of
the checkerboard states is shown in Fig. 3. The energetics of the
checkerboard state has been studied extensively in Ref. \cite{CHEN2004},
both numerically and analytically. In general, at the magic doping
fraction $x=(2m+1)/2^n$, the charge unit-cell is $2^{(n+1)/2}a \times
2^{(n+1)/2}a$, pointing along the original Cu-O bond direction when
$n$ is odd, and along the diagonal direction when $n$ is even. This
theory relies on mapping the original fermionic degrees of freedom into
effective bosonic degrees of freedom \cite{CHEN2003}; such mapping may
be justified in the underdoped and optimally doped regimes, but fails in
heavily doped samples. Therefore, while the bosonic theory predicts all
magic doping fractions at $x=(2m+1)/2^n$, one can only expect the
effective theory to be valid for $x<1/4=25\%$. Also, it is generally
expected that the charge ordering tendencies are stronger at higher
levels of the hierarchy, with smaller $n$.

\begin{figure}
\includegraphics[clip,width=8.5cm]{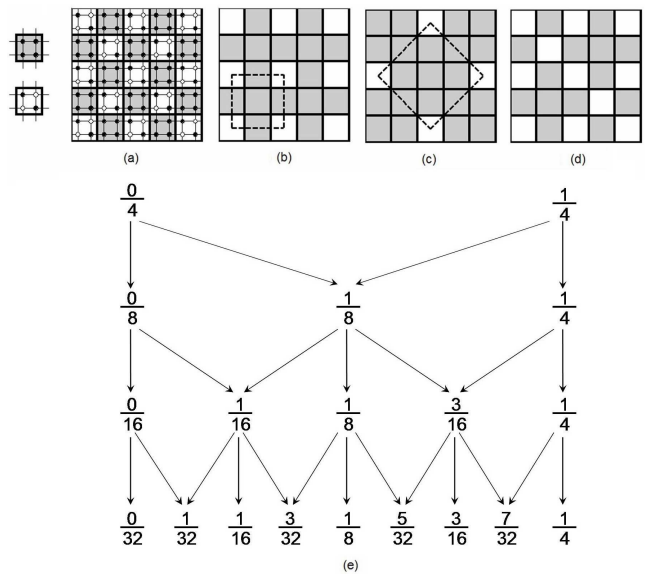}
\caption{Hierarchical construction of the checkerboard-type
ordering of the hole pairs at ``magic" doping fractions
$(2m+1)/2^n$, where $m$ and $n$ are integers. Following the
construction of Ref. \cite{CHEN2003}, the original CuO$_2$ lattice
is grouped into non-overlapping plaquettes, which can be
represented by the squares on a checkerboard. A checkerboard can
be alternately colored black and white; in our case, each black
square contains four sites and no holes, while each white square
contains four sites and two holes in the form of a Cooper pair.
Such a state has hole doping density $x=1/4$ (a), as
represented at the highest level of the hierarchy (e).
(Electrons are denoted by black dots, and holes are denoted by
open dots; since we only address the issue of charge ordering
here, the spin of the electron is not explicitly indicated).
At the next level of the hierarchy, consider the lattice of white
squares only, and alternately color half of them black. Such a
state has hole doping density $x=1/8$ (b). At one further
level down in the hierarchy, one can either consider the lattice
of the white squares, and alternately color half of them black,
thus obtaining a state with $x=1/16$ (c), or one can
consider the lattice of the newly colored black squares and
alternately color half of them white, thus obtaining a state with
$x=3/16$ (d). This hierarchy construction can be obviously
iterated {\it ad infinitum}, generating a binary tree of magic
doping fractions as shown here in (e).}
\end{figure}

Both the 1D stripe model and the 2D checkerboard model adequately
explain the dominant magic doping fraction at $x=1/8$, but they predict
different sets of magic doping fractions, at which the system is
expected to develop charge ordering tendencies. In this regard, our
extensive data set on the doping dependence, if it indeed reflects a
charge ordering tendency, contains sufficient accuracy to distinguish
between the simple 1D stripe model and the 2D checkerboard model
discussed above. The simple stripe model predicts commensurate effect at
magic doping fractions 1/4, 1/6, 1/8, 1/10, 1/12, 1/14, 1/16, which
should be either equally strong or vary monotonically in strength;
therefore, the absence of any commensurability effects at $x=1/6$ and
$x=1/10$ in our data or any other previous experiments is puzzling in
the simple stripe model. Although the stripe structure can yield a
complex ``devil's staircase" of commensurate dopings in nickelates
\cite{WOCHNER1998}, the magic fractions suggested here in a cuprate
superconductor would be a challenge to the stripe picture. On the other
hand, the suggested series (1/16, 3/32, 1/8, and 3/16) agrees
surprisingly well with the magic doping fractions predicted from the
checkerboard model discussed above, up to the level $n=5$. At this
level, the absence of the $1/32$ fraction is understandable, since the
hole-pair lattice at this doping fraction would be very dilute and
therefore disordered. More notable is the absence of the $5/32$
fraction, for which we do not have an adequate explanation at this
moment. In passing, let us briefly mention the Wigner crystal of single
holes, which has the charge unit-cell $2^{n/2}a \times 2^{n/2}a$ at
$x=1/2^n$. The main difference between the pair checkerboard and the
hole checkerboard is not only the size of the charge unit-cell but also
their orientations, which are $45^{\circ}$ with respect to each other;
thus, the two should be easily discernible in experiments at a given
doping.

Neutron scattering on the LSCO-based materials have provided an
extensive set of data on the {\it spin} order. We note, however, that
the nature of the spin order may not be directly related to the nature
of the charge order reflected in the transport properties, particularly
in the superconducting doping regime of LSCO where the spin order is
mostly dynamic \cite{YAMADA1998}. If the magnetic incommensurability
observed by the neutron scattering is part of some dynamic dispersing
mode \cite{CHRISTENSEN2004,NORMAN2000,HAYDEN2004}, it is natural that
the incommensurability does not represent the unit-cell of the incipient
charge order. Furthermore, at $x$ = 0.02, the magnetic neutron
scattering found static and unidirectional spin stripes that are no less
than 30 unit-cell apart (magnetic incommensurability $\delta$ is 0.016)
\cite{MATSUDA2000}, which would cause a large resistivity anisotropy if
the charges are conforming to the 1D spin stripes. However, transport
measurements on untwinned single crystals have found only a factor of
1.5 resistivity anisotropy between the ``longitudinal" and ``transverse"
directions \cite{ANDO2002}; this is rather difficult to understand
without invoking some 2D character (which can, for example, be coming
from a nematic stripe order \cite{Kivelson1998}) in the charge system.
In passing, we note that our transport measurements of LSCO have found
evidence for charge self-organization \cite{ANDO2001,ANDO2003} and
modest one-dimensionality \cite{ANDO2002,DUMM2003} only in the lightly
doped regime ($x \le 0.05$) of this compound, while the present study
suggests the existence of the checkerboard order only in the
superconducting doping regime ($x \ge 0.06$). We also note that the
measurements of the in-plane resistivity anisotropy become impractical
for $x \ge 0.06$, because the structural transition temperature (below
which the system becomes orthorhombic) comes close to or below the room
temperature at these dopings, making it difficult to prepare untwinned
samples.

There are only a few direct experimental observations of charge order in
the LSCO-based materials. Neutron scattering on the
La$_{1.48}$Nd$_{0.4}$Sr$_{0.12}$CuO$_4$ (LNSCO) compound at $x=1/8$
reveals elastic charge order peaks \cite{TRANQUADA1996} which can be
interpreted either as orthogonally intersecting stripes on alternating
planes or in the same plane. The latter case would also be consistent
with the 2D pair checkerboard pattern with the charge unit-cell $4a
\times 4a$. While it is fair to mention that the details of the charge
order in LNSCO are more consistent with the 1D stripe picture
\cite{ZIMMERMANN1998}, such a structure may well be a result of the 1D
modulation arising from the explicit symmetry breaking in the
low-temperature tetragonal (LTT) phase of this particular material, and
therefore may not be representative of the charge order in Nd-free LSCO.

In view of the intriguing agreement of the present transport data with
the hole pair checkerboard model, it would be desirable to
systematically perform direct measurements of the charge order in the
LSCO-based materials by some means. If the proposed checkerboard states
are indeed realized in LSCO, the orientation of the charge unit cell
should be along the Cu-O bond direction at $x=1/8$ (as is the case in
BSCCO or NCCOC), while it should be along the diagonal direction near
$x=1/16$; it would be definitive if this rotation of the charge unit
cell upon changing $x$ is confirmed by a direct means. Also, since some
fractions suggested in the present measurements are relatively weak, it
would be highly desirable to carry out these systematic transport
measurements under high magnetic fields or under high pressure, where
one would expect the competing order to be enhanced and the magic doping
fractions to be more pronounced.

We would like to thank P. W. Anderson, S. A. Kivelson, J. M. Tranquada,
A. Yazdani and Z. X. Zhao for helpful discussions. This work is
supported by the Grant-in-Aid for Science provided by the Japanese
Society for the Promotion of Science, the NSF under grant numbers
DMR-0342832, and the US Department of Energy, Office of Basic Energy
Sciences, under contract DE-AC03-76SF00515.


\bibliography{MagicNumber}

\end{document}